\documentclass[amsmath,amssymb,aps,prl,reprint,floatfix,superscriptaddress]{revtex4-2}

\usepackage{graphicx}
\usepackage{dcolumn}
\usepackage{bm}
\usepackage[version=4]{mhchem}
\usepackage{xcolor}
\usepackage[breaklinks,colorlinks,linkcolor={blue}, %
            citecolor={blue},urlcolor={blue}]{hyperref}
\begin{document}

\title{High-throughput identification of spin-photon interfaces in silicon}

\author{Yihuang Xiong} 
\affiliation{Thayer School of Engineering, Dartmouth College, Hanover, New Hampshire 03755, USA}

\author{C\'eline Bourgois}
\affiliation{Institute of Condensed Matter and Nanosciences (IMCN), Universit\'{e} Catholique de Louvain, Chemin des \'{E}toiles 8, B-1348 Louvain-la-Neuve, Belgium}
\affiliation{Thayer School of Engineering, Dartmouth College, Hanover, New Hampshire 03755, USA}

\author{Natalya Sheremetyeva}
\affiliation{Thayer School of Engineering, Dartmouth College, Hanover, New Hampshire 03755, USA}

\author{Wei Chen}
\affiliation{Institute of Condensed Matter and Nanosciences (IMCN), Universit\'{e} Catholique de Louvain, Chemin des \'{E}toiles 8, B-1348 Louvain-la-Neuve, Belgium}

\author{Diana Dahliah}
\affiliation{Institute of Condensed Matter and Nanosciences (IMCN), Universit\'{e} Catholique de Louvain, Chemin des \'{E}toiles 8, B-1348 Louvain-la-Neuve, Belgium}

\author{Hanbin Song}
\affiliation{Department of Electrical Engineering and Computer Sciences, University of California, Berkeley, California 94720, USA}

\author{Sin\'ead M.\ Griffin}
\affiliation{Materials Sciences Division, Lawrence Berkeley National Laboratory, Berkeley, California 94720, USA}
\affiliation{Molecular Foundry Division, Lawrence Berkeley National Laboratory, Berkeley, California 94720, USA}

\author{Alp Sipahigil}
\affiliation{Department of Electrical Engineering and Computer Sciences, University of California, Berkeley, California 94720, USA}
\affiliation{Department of Physics, University of California, Berkeley, California 94720, USA}
\affiliation{Materials Sciences Division, Lawrence Berkeley National Laboratory, Berkeley, California 94720, USA}

\author{Geoffroy Hautier} 
\affiliation{Thayer School of Engineering, Dartmouth College, Hanover, New Hampshire 03755, USA}
\date{\today}

\begin{abstract}
Color centers in host semiconductors are prime candidates for spin-photon interfaces that would enable numerous quantum applications. The discovery of an optimal spin-photon interface in silicon would move quantum information technologies towards a mature semiconductor technology. However, the space of possible charged defects in a host is very large, making the identification of promising quantum defects from experiments only extremely challenging. Here, we use high-throughput first principles computational screening to identify spin-photon interfaces among more than 1000 substitutional and interstitial charged defects in silicon. We evaluate the most promising defects by considering  their optical properties, spin multiplicity, and formation energies. The use of a single-shot hybrid functional approach is critical in enabling the screening of a large number of defects with a reasonable accuracy in the calculated optical and electronic properties. We identify three new promising spin-photon interface as potential bright emitters in the telecom band: $\rm Ti_{i}^{+}$, $\rm Fe_{i}^{0}$, and $\rm Ru_{i}^{0}$. These candidates are excited through defect-bound excitons, stressing the importance of considering these type of defects in silicon if operations in the telecom band is targeted. Our work paves the way to further large scale computational screening for quantum defects in silicon and other hosts. 
\end{abstract}

\maketitle

\section{Introduction}

Point defects have become central to many quantum technologies from sensing to communication and computing. A color center in a semiconducting host can act as an artificial atom and be used to harness quantum effects~\cite{Wolfowicz2021, Yan2021,  Zhang2020}. For instance, these quantum defects can be used as spin-photon interfaces where their electronic spin are initialized, controlled and read by photons. Important spin-photon interfaces have been studied in diamond such as the \textit{NV} center or the silicon divacancy and can constitute multi-qubit quantum registers based on electron and nuclear spins that are optically initialized, measured and entangled over long distances~\cite{Hensen2015Nature,Bradley2019PRX, Stas2022, Lukin2020, Atature2018}. Finding a perfect spin-photon interface is challenging as the color center needs to exhibit a series of properties from a bright emission in an adequate wavelength, to spin multiplicity and optical as well as spin coherence~\cite{Bassett2019}. While most of the work on quantum defects has been performed using wide band gap hosts such as diamond and silicon carbide, moving to silicon would have many advantages for integration and nanofabrication. Despite silicon's smaller bandgap, recent promising results using carbon-based complex defects produced by ion-implantation such as the \textit{T} or \textit{G} centers have opened up silicon as a promising defect host~\cite{Bergeron2020, MacQuarrie2021, Redjem2020, Durand2021, Baron2022, Morse2017, Yan2021,Komza2022}. These color centers were discovered serendipitously before the advent of quantum information science when studying ion-implantation of typical contaminants in silicon. However, since the chemical space of possible defects is very large, it has led to speculation whether a yet-to-be discovered quantum defect could act as high performance spin-photon interface in silicon.

First principles techniques are commonly used to understand and propose new quantum defects~\cite{Dreyer2018, Alkauskas2016, Smart2021, Gali2019, Tsai2022}. These atomistic modeling techniques have matured enough that they can now be used on a large scale to search for materials with specific properties in a high-throughput fashion~\cite{Jain2013, Curtarolo2013, hautier2019}. However, screening efforts have so far mainly focused on bulk properties with only a few attempts at point defect related properties~\cite{Bertoldo2022, Mannodi-Kanakkithodi2022, Mannodi-Kanakkithodi2022a, Dahliah2021, Emery2017, Kumagai2021, Davidsson2022}. Recent developments in the automation of charged defect computations offer the opportunity to fill this gap but large-scale defect computations still face important challenges, especially balancing accuracy with computational cost~\cite{Broberg2018, Goyal2017, Huang2022DASP, Davidsson2021}. The least computationally expensive approach, the semilocal density functional theory (DFT), often fails to provide reliable opto-electronic properties, while more accurate approaches such as hybrid functionals are more computationally demanding and can only be used on small sets of possible defects limiting the screening space~\cite{Lee2022, Smart2021, Broberg2023}. Additionally, most of the efforts in predicting new quantum defects have focused on mimicking the \textit{NV} center in diamond and assumed that only defects presenting at least two single-particle energy levels well within the band gap are of interest \cite{Varley2016, Tsai2022, Lee2022,Xiong2023arXiv}. This might not be a viable approach in silicon where color centers are commonly excited through the formation of defect-bound excitons and thus involve delocalized band-like states~\cite{Davies1989}. In fact, the most studied spin-photon interface candidates in silicon (i.e., \textit{T} center and $\rm Se_{Si}^{+}$) do form defect-bound excitons,  highlighting the need to consider them in high-throughput screening approaches~\cite{Bergeron2020, Morse2017, Dhaliah2022,Xiong2023arXiv}.

Here, we report on a high-throughput computational search of quantum defects for high performance spin-photon interfaces in silicon. Searching for point defects made of an unique foreign element addition (among 56 elements) in silicon either as substitutional or interstitial defect, we generate a computational database of around a thousand charged defects. We identify unique color centers combining spin multiplicity and a bright emission in the short wave infrared (SWIR) or telecom-band. We use a combination of techniques based on semilocal DFT and hybrid functionals that balance predictive accuracy with computational cost. Importantly, we refrain from updating the semilocal DFT wavefunctions in the hybrid functional calculations. This single-shot hybrid functional greatly reduces the computational cost while retaining the overall accuracy of the fully self-consistent version for single-particle energies. We perform our search allowing optical transitions involving host-like states through defect-bound excitons and find that this is essential for defect candidates active in the near-infrared in silicon. After presenting our spin-photon interface candidates, we provide the fundamental chemical and structural reasons for their emergence from our screening and discuss inherent challenges for quantum defects in silicon.

\section{Results}
A spin-photon interface is a color center where the spin state can be controlled, initialized, and read out by light. We focus here on a minimal set of necessary requirements for such a quantum defect. First, the defect ground state needs to have a non-zero total electronic spin (i.e., a spin multiplicity different than one). This electronic spin state is controlled by light excitation and emission through photoluminescence. This requires the defect to have excited states accessible optically in a technologically relevant wavelength. Here, we target the SWIR wavelength compatible with silicon photonics of at most 1700~nm (i.e., at least 700~meV or 5646~cm$^{-1}$), amenable to fiber-optic transmission and standard infrared detectors. Optical emissions need to be allowed and bright implying that the radiative lifetime of the excited state should be short. Additional requirements are for the defect emission to occur mainly through the zero-phonon line (ZPL) peak and to have a minimal amount of phonon-sideband emission (i.e., a low Huang-Rhys factor). Note that these requirements are necessary but not sufficient. For example, the optical and spin coherence and the ease of control and nanofabrication are
also important, but the primary requirements we have outlined rely on properties which can be assessed on a large scale by first-principles computations. Notably, the results we will present here show that our limited screening criteria are already quite restrictive in selecting defect candidates.

\begin{figure}[t]
 	\includegraphics[width=0.45\textwidth]{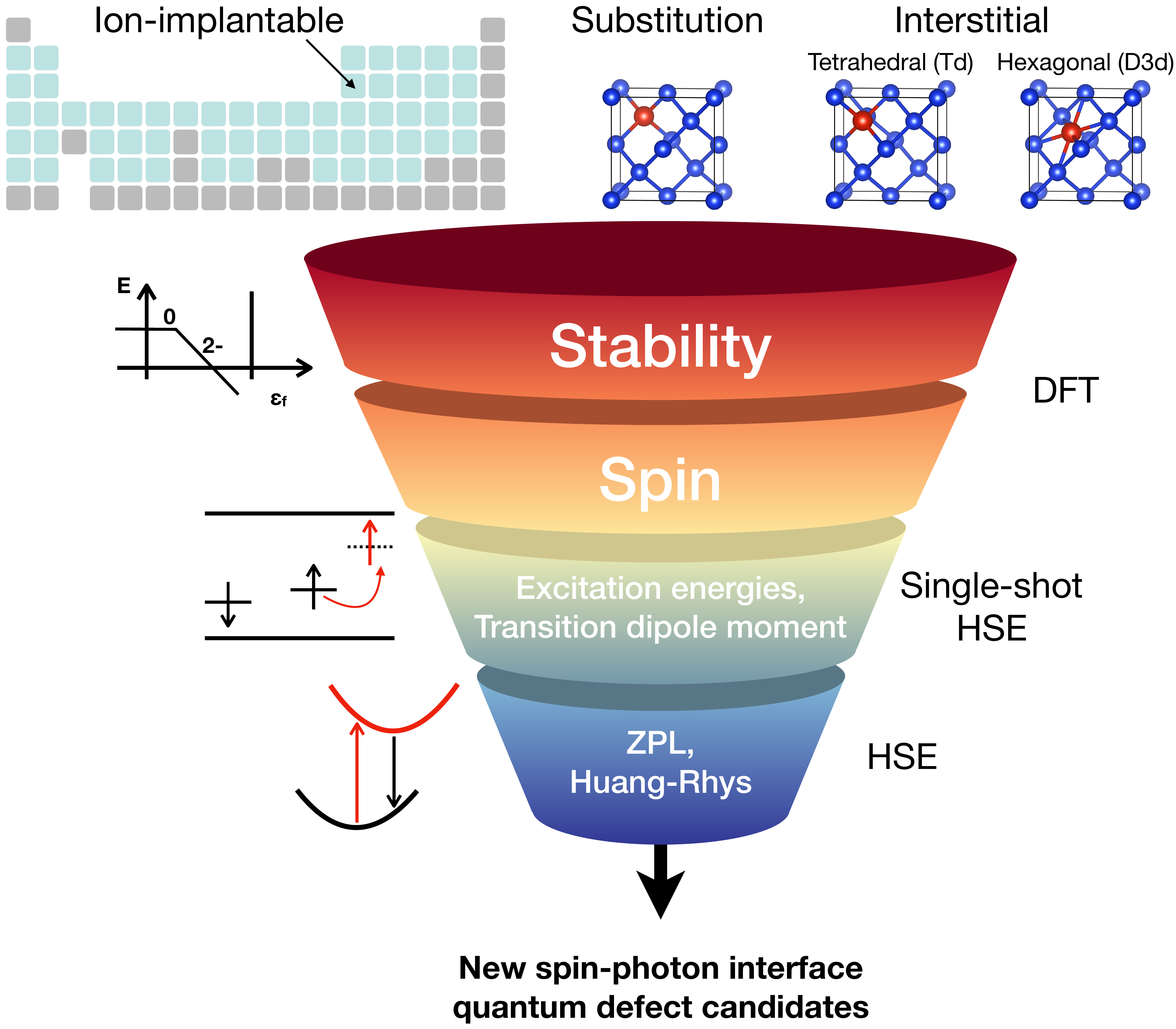}    
 	\caption{\label{Fig.1 screening tiers}Workflow for screening quantum defects in silicon for spin-photon interface. Three types of simple defects were considered in this work, including substitutional, tetrahedral interstitial ($T_d$), and hexagonal interstitial ($D_{3d}$). 
 	}
\end{figure}

We consider three `simple' defects (in contrast to complex defects) possible in silicon: substitutional, tetrahedral interstitial, and hexagonal interstitial defects (see Fig. \ref{Fig.1 screening tiers}) using a set of 56 elements identified as easily ion-implantable (see Methods). Our approach relies on a tiered screening approach outlined in Fig. ~\ref{Fig.1 screening tiers} to detect defects with high brightness, appropriate ZPL energy, and low Huang-Rhys factor. We compute these three types of defects in a range of charge states (based on common oxidation states of the elements) at the DFT level. This leads to a total of 1042 defect computations performed entirely automatically. Certain charged defects are not stable for any Fermi level, we do not expect these to be achievable in experiments even using non-equilibrium techniques such as ion-implantation. Our first screening is to only consider defects that are stable for any range of Fermi level (617 charged defects remain). For each of these stable charged defects, we use the single-particle Kohn-Sham (KS) eigenvalues to identify the possible optical transitions between occupied and unoccupied states within a spin channel. However, neither the band gap nor the single-particle defect energy level is properly described by semilocal DFT \cite{Perdew1981, Yang2008}. Higher levels of theory, such as hybrid functionals (e.g., the Heyd-Scuseria-Ernzheroff (HSE) functional~\cite{Heyd2003}) and the 
 $GW$ Approximation, can provide more accurate single-particle energy levels but at a significantly higher computational cost. Here, we use the `single-shot hybrid functional' approach (HSE$_0$ hereafter) where the DFT wavefunctions are applied to the HSE Hamiltonian without further iteration (see methods).  This single-shot HSE approach provides a significant improvement to DFT single-particle energies for only a marginal increase of computational cost (around 5\% of computational overhead from DFT). The technique is closely related to the single-shot $GW$ approach (i.e., $G_0W_0$) for which the quasi-particle energies are computed on top of the DFT wavefunctions. The HSE$_0$ approach offers an excellent compromise between accuracy and computational cost as outlined in SI (see Fig. S1) and has been used successfully in the past on defects but never in a high-throughput context~\cite{Alkauskas2007}. 
 In addition, for each optical transition between an occupied and an unoccupied single-particle energy level, we compute the transition dipole moment (TDM) $\boldsymbol{\mu} \propto \langle {\psi}_{i}| \boldsymbol{r}| {\psi}_{f} \rangle$ at the DFT level to assess the radiative lifetime of the excited state. A high transition dipole moment is an important requirement which indicates strong emission brightness. We also used the HSE$_0$ corrected single-particle energies to assess the total spin of the defect and we focus on defects in a non-singlet state (total spin $S \neq 0$). This first screening identifies promising defects that we follow up with full ionic and electronic relaxation at the HSE level to obtain the ZPL and Huang-Rhys factors. More details on the high-throughput screening are available in the methods section.

\begin{figure*}[ht]
 	\includegraphics[width=\textwidth]{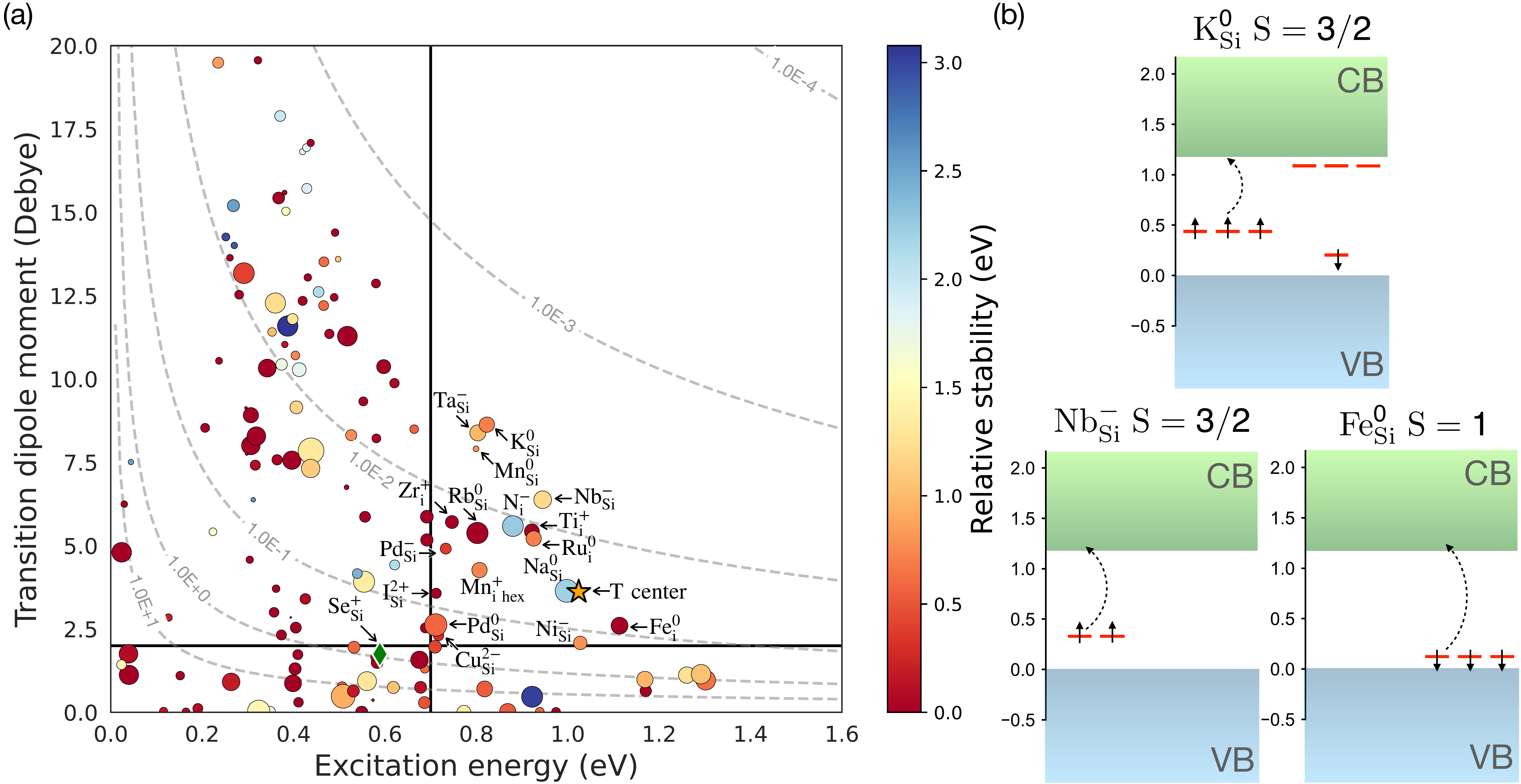}    
 	\caption{\label{Fig.2 TDM_vs_E}(a) Transition dipole moment vs.\ single-particle excitation energy at the single-shot HSE (HSE$_0$) level for the stable defects with non-singlet spin multiplicity. The color of the circles indicates how stable the structure of the defect is as an interstitial defect relative to a substitutional one, while the size represents the stability window of the charged defect within the band gap. The isolines mark the radiative lifetime in $\mu$s. For comparison, we highlight the \textit{T} center (star) and $\rm Se^{+}_{Si}$ (diamond). $\rm Co^{2-}_{Si}$ is removed from the figure due to nonphysically large shift of the KS levels from DFT to HSE$_0$. (b) Single-particle excitation energies at the HSE$_0$ level of theory for a series of candidates: $\rm K_{Si}^{0}$, $\rm Nb_{Si}^{–}$, $\rm Fe_{i}^{0}$. The defect levels are shown in red.}
\end{figure*}

Fig. \ref{Fig.2 TDM_vs_E}(a) plots the TDM versus the lowest optical transition energy for all stable non-singlet charged defects in our database computed at the HSE$_0$ level. The size and color of the data points relate to their charge stability (the extent of the Fermi level within the band gap where the charge state is stable) and the stability of the defect structure (interstitial versus substitutional), respectively. The plots including singlet ($S=0$) charged defects are available in Fig.~S2. For comparison, we highlight in Fig.~\ref{Fig.2 TDM_vs_E} two defects previously considered as spin-photon interfaces in silicon: the \textit{T} center and the $\rm  Se_{Si}^{+}$ defect. The \textit{T} center is not in our dataset as it is a complex defect made of carbon and hydrogen. The $\rm Se_{Si}^{+}$ and \textit{T} center show attractive TDMs (1.75 and 3 D at the DFT level, respectively). $\rm Se_{Si}^{+}$ operates however at a low excitation energy (0.59~eV) while the \textit{T} center is closer to our target (1.024~eV). Our results agree with the experimental knowledge on these defects~\cite{Bergeron2020, Morse2017}.
 A general anticorrelation is observed between TDM and transition energy as larger excitation energies lead to lower TDMs and hence dimmer optical transitions. Defects combining a high TDM and a large emission energy are therefore rare. Setting a threshold of 700~meV excitation energy and a TDM of at least 2~D, we find 19 defects among our 617 stable charged defects (see Table~S1). As an illustration of the type of defects selected, Fig.~\ref{Fig.2 TDM_vs_E}b shows the single-particle energies for three selected defects: $\rm K_{Si}^{0}$, $\rm  Fe_{i}^{0}$, and $\rm  Nb_{Si}^{-}$. It is envisaged from the single-particle energies that the transition can take place from a localized defect state within the gap to a delocalized host-like state in the form of a bound exciton. In fact, the vast majority of the 19 defects candidates are excited as defect-bound excitons.

For these 19 candidates, we refine our screening and run full ionic and electronic relaxations with the fully self-consistent HSE calculations. The single-particle levels with respect to the band edges are plotted in Fig.~S3. Only 7 defects combine a high TDM and a high transition energy (i.e., single-particle transition energy over $800$~meV): $\rm Fe_{i}^{0}$, $\rm Ti_{i}^{+}$, $\rm Zr_{i}^{+}$, $\rm Ru_{i}^{0}$, $\rm Nb_{Si}^{-}$, $\rm Ni_{Si}^{-}$, and $\rm Na_{Si}^{0}$. We note that these transition energies do not take into account electronic and ionic relaxation of the excited state and tend to overestimate the ZPL (typically by at least 100 meV).

For many other defects, the electronic structure obtained with HSE$_0$ is close to the full HSE results but the excitation energies remain too low ($<$ 800~meV) to make them likely to offer an appealing ZPL: $\rm Mn_{i\ hex}^{0}$, $\rm Co_{Si}^{2-}$, $\rm Cu_{Si}^{2-}$, $\rm I_{Si}^{2+}$, $\rm K_{Si}^{0}$, $\rm Rb_{Si}^{0}$, $\rm Pd_{Si}^{0}$, $\rm Pd_{Si}^{-}$, $\rm Ta_{Si}^{-}$. Importantly, we find that a reorganization of energy level happens when full HSE is performed for a few defects: $\rm Co_{i}^{+}$, $\rm Mn_{Si}^{0}$ and $\rm N_{i}^{-}$. This stresses the importance of using a tiered screening approach and validating our single-shot HSE computation with further full HSE computations.

We compute the ZPL and formation energy versus Fermi level for the 7 best candidates identified above. The ZPL was computed by relaxing the structure and imposing the occupation of the lowest unoccupied state following the constrained-HSE approach (see Methods). The results are presented in Table \ref{Tab: non-singlet defects} and Fig.~S4. All the ZPLs are higher than 700~meV except for $\rm Zr_{i}^{+}$ which sits at 666~meV. The full HSE computations confirm that these defects are all excited through an electronic transition between a localized state and a host-like state forming a defect-bound exciton. The excited electron is promoted either from a localized defect state to a conduction-band state (donor-bound exciton) or from a valence-band state to a localized defect state (acceptor-bound exciton). For instance, the $\rm Fe_{i}^{0}$ donor-bound exciton is formed by a delocalized electron attracted by the potential induced by a $\rm Fe_{i}^{+}$ charged defect. 

\begin{table*}[ht!]
    \caption{\label{Tab: non-singlet defects}Kohn-Sham level difference ($\Delta $KS), zero-phonon line (ZPL), transition dipole moment (TDM), the relevant charge transition level (CTL) with respect to $E_{\rm VBM}$, bound exciton stability (BES), and the nature of excitation of the 7 best simple defect candidates that were screened. All reported values are computed in 512-atoms cell and at HSE level.}
    \begin{ruledtabular}
    \begin{tabular}{c c c c c c c c}
Defect & $\Delta$KS (eV) & Total spin & ZPL (meV) & TDM (Debye) & CTL (eV) & BES (meV) & Nature of excitation\\ \hline
 $\rm Ti_{i}^{+}$& 1.04 & 3/2 & 939 & 1.86 & 0.130 (+2/+1) & 45 & Donor\\
 $\rm Zr_{i}^{+}$& 0.81 & 3/2 & 666 & 3.01 & 0.263 (+2/+1) & 666 & Donor\\
 $\rm Fe_{i}^{0}$& 0.98 & 3/2 & 903 & 1.59 & 0.291 (+2/0) & 211 & Donor\\
 $\rm Ru_{i}^{0}$& 1.10 & 1 & 791 & 1.15 & 1.10 (0/$-1$) & 311 & Acceptor\\
 $\rm Na_{Si}^{0}$& 1.03 & 3/2 & 973 & 0.93 & 0.363 (+1/0) & $-222$ & Donor\\
 $\rm Nb_{Si}^{–}$& 0.97 & 1 & 908 & 1.09 & 0.642 ($-1$/$-2$) & $-436$ & Donor\\
 $\rm Ni_{Si}^{-}$& 0.92 & 3/2 & 870 & 1.01 & 0.592 ($-1$/$-2$) & $-348$ & Donor
\end{tabular}
\end{ruledtabular}
\end{table*}

When considering defects excited through bound excitons, we need to keep in mind a competing process that could destabilize the exciton. The electron or hole in the bound exciton perturbed host-like state could be released and freed as a free electron or hole. This corresponds to the breaking of the exciton. For instance, in the case of iron, the bound exciton composed of an electron bound to $\rm Fe_{i}^{+}$ could break and lead to $\rm Fe_{i}^{+}$ and a free electron. How favorable this ionization process is can be computed comparing the ZPL to the relevant charge transition level (obtained from defect formation energy plots). This provides an estimate of the exciton binding energy. We have computed the exciton binding energy for our 7 defects candidates as reported in Table~\ref{Tab: non-singlet defects}. For all defects but $\rm Nb_{Si}^{-}$, $\rm Ni_{Si}^{-}$ and $\rm Na_{Si}^{0}$ the exciton is stable (at low temperature) given the positive binding energy. The instability of the negatively charged donor bound exciton (i.e., $\rm Nb_{Si}^{-}$ and $\rm Ni_{Si}^{-}$) is not surprising. Here, the exciton would be formed by an electron bound to a neutral defect ($\rm Nb_{Si}^{0}$ or $\rm Ni_{Si}^{0}$). The absence of electrostatic energy binding the electron to the neutral defect explains the instability of the exciton. By contrast, neutral defects such as $\rm Fe_{i}^{0}$ or positive defects such as $\rm Ti_{i}^{+}$ can stabilize a donor bound exciton through the attractive Coulomb interaction between the bound electron and the positively charged defect (either +1 or +2).

\begin{figure}[t]
 	\includegraphics[width=0.5\textwidth]{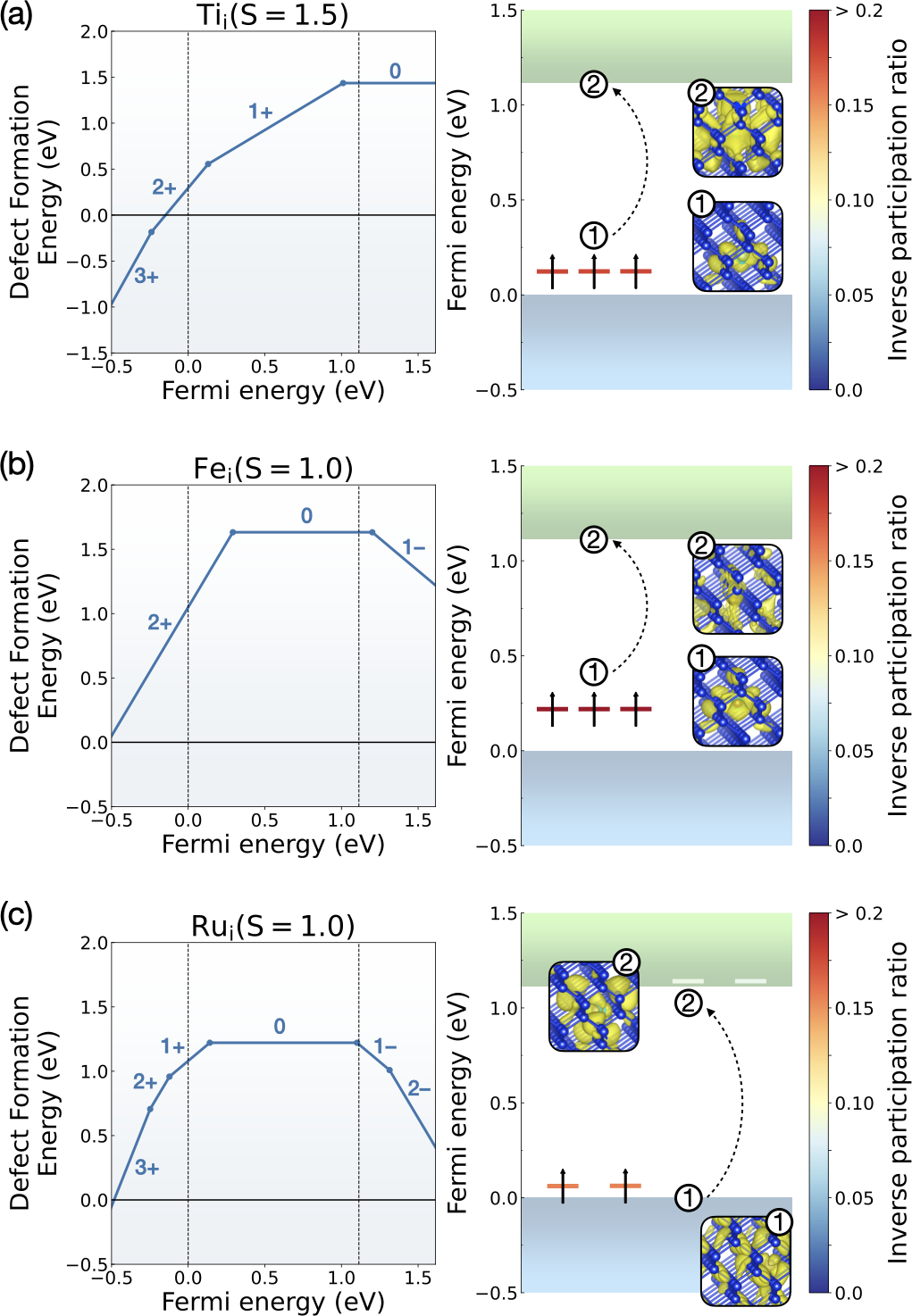}
 	\caption{\label{Fig.3 three candidates}The defect formation energies, single-particle defect level diagrams, and the charge density for the relevant bands for $\rm Ti_{i}^{+}$, $\rm Fe_{i}^{0}$, $\rm Ru_{i}^{0}$. For simplicity, we only plot the defect levels in the gap. The localization of the defect levels are represented by the inverse participation ratio. The charge density for the localized orbitals are shown with an isocontour value of 0.0005 $\AA^{–3}$, and the delocalized orbitals are shown with a value of 0.00008 $\AA^{–3}$.}
\end{figure}

After this extensive screening we have found three defect candidates of interest for spin-photon interface: $\rm Ru_{i}^{0}$, $\rm Fe_{i}^{0}$, and $\rm Ti_{i}^{+}$. Fig.~\ref{Fig.3 three candidates} shows the formation energy vs Fermi level and the single-particle levels for all our candidates. The defect-bound exciton nature of the excited state of these defects is again noticeable. The range of stability of the required charged state within the Fermi level is large for all defects indicating that they should be relatively easily formed in the right charged state. The $\rm Ti_{i}^{+}$ will be favored in insulating silicon avoiding highly n- or p-type doped host. $\rm Ru_{i}^{0}$ and $\rm Fe_{i}^{0}$ will form in either insulating or n-type doped silicon. 

The TDM of these defects is technologically attractive from 1.86 to 1.15 D, higher than that of the \textit{T} center (0.43 D with the same level of theory). The radiative lifetime estimated from these TDM are between 0.40 and 0.89 $\mu$, which are lower than the \textit{T} center (5 $\mu$s) but still not as low as that of the \textit{NV} center in diamond (around a few nanoseconds)~\cite{Batalov2008, Goldman2015}.

The $\rm Ti_{i}^{+}$, $\rm Fe_{i}^{0}$ interstitial defects are energetically favored versus their substitutional counterparts and should be relatively easy to form (see Fig. S5). For ruthenium, while the interstitial defect is of interest, the substitutional configuration is energetically favored. This defect could however potentially be made interstitial through ion-implantation which is known to be able to produce non-equilibrium defect configurations.

We have also computed the full PL spectra including phonon side-bands and the Huang-Rhys factor for $\rm Ti_{i}^{+}$, $\rm Fe_{i}^{0}$ and $\rm Ru_{i}^{0}$ (see Fig. S6a). The phonon side-bands are remarkably weak for all defects. The Huang-Rhys factor measures the average number of phonons participating to the emission, and is computed to be 0.18, 0.32 and 0.36 respectively. These are much lower than the computed Huang-Rhys factor for the \textit{NV} center which is 3.76~\cite{Razinkovas2021,Alkauskas2014}. A lower Huang-Rhys factor will lead to lower losses through the phonon sideband during emission and is advantageous technologically. We hypothesize that simple interstitial or substitutional defects might have lower Huang-Rhys factors than complex defects due to their lower number of internal degrees of freedom. This might be an inherent advantage of simple defects versus complex defects. The previously known $\rm Se_{Si}^{+}$ substitutional quantum defect shows a calculated Huang-Rhys factor of about 0.70, see Fig. S6b.

When comparing our results to experimental literature, we note that it is not always straightforward to link a measured line or signal in PL or deep level transient spectroscopy (DLTS) to a specific defect. These measurements are typically made on an ensemble of defects and might include uncontrolled complexes. Moreover, optical data (PL or absorption) is sparser than electronic data on silicon defects. Titanium and iron show DLTS data in agreement with our charge transition level computations. For iron, the computed charge transition levels within HSE are for the +2/0 level at 0.29 eV above the valence-band maximum (VBM) while DLTS experiments report a transition level at 0.38 eV~\cite{Graff2000}. This is within the range of error seen in HSE. Optical data points out to an optical transition related to the excitation of electrons in the +2/0 to the conduction band around 750 and 780~meV~\cite{Istratov1999}. This is in fair agreement with our computed ZPL at 903 meV and is consistent with the 100~meV difference between HSE and experiment for the +2/0 charge transition level. Moreover, the bound excitonic nature of this optical transition has been reported experimentally and agrees with our computations~\cite{Thilderkvist1993}. Remarkably, the PL lines exhibit a Zeeman effect in agreement with our theoretical prediction of a non-singlet ground state and confirming the potential for $\rm Fe_{i}^{0}$ as a spin-photon interface~\cite{Thilderkvist1998}. We note that there are some discussions in the literature about the assignment of this line to Fe interstitial in view of isotope shift data~\cite{Istratov1999, Schlesinger1983}. For titanium, we have a +2/+1 level at 0.13 eV according to HSE which compares reasonably well to the electronic transition measured at 0.28 eV experimentally~\cite{Graff2000}. We however did not find reliable experimental optical absorption or PL data for titanium. For both iron and titanium defects, we see that despite being common transition metals the optical properties of these defects are not always clear. Many of these previous experimental results have important uncertainty that we hope the recent experimental developments in high resolution spectroscopy using potentially $\rm ^{28}Si$ enriched samples and measurement of individual defects in silicon could clarify~\cite{Zhou1991, Chen1980, Steger2011, Redjem2020, Bergeron2020, Baron2022}. A direct comparison between experiment and our results on ruthenium interstitial is more difficult as we expect the substitutional defect $\rm Ru_{Si}$ to be the most stable (see Fig.~S5). We note that the limited DLTS data on ruthenium in silicon is even not consistent with our computations for the substitutional defect. Our HSE$_0$ computations do not show any defect level in the band gap while the DLTS results indicate a series of deep levels (see SI, Fig.~S7(c))~\cite{Chen1980}. These experimentally measured defect levels are likely due to complexes. We note that in SiC, a PL signature attributed to interstitial ruthenium has been observed despite its higher formation energy compared to the substitutional configuration as in silicon \cite{Kleppinger2019}. This suggests that the ruthenium interstitial defect could be produced despite its energetic preference for a substitutional position.

The number of defect candidate is strikingly limited, i.e., only 3 from a pool of 1024 charged defects. This stresses the challenge in discovering high performance spin-photon interfaces in silicon and motivates the use of computational screening techniques. Our work calls for experimental efforts towards the realization of these three defect candidates especially $\rm Fe_{i}^{0}$ and $\rm Ti_{i}^{1+}$. Additionally, further computational efforts using techniques that are not yet high-throughput would further clarify the properties of these promising defects~\cite{Onizhuk2021,Huang.PRX.2022,Ma.JCTC.2021,Jin.npj.2022}.

\section{Discussion}

Using high-throughput computational screening, we identified a series of potential spin-photon interface candidates in silicon. Beyond the discovery of specific new quantum defects, our dataset can be used to discuss the general trends and limits in properties for simple defects in silicon and to further understand the inherent chemical and structural reasons leading to high performance spin-photon interfaces.

The symmetry of the point defect has important implications on its spectral properties. The absence of an electric dipole in centrosymmetric defects reduces their coupling to electric fields, making them immune to spectral diffusion. This is an important technological advantage that motivated the interest in the silicon divacancy in diamond~\cite{Udvarhelyi2019,Jahnke2015}. Among the three types of simple defects considered in our study, both the interstitial tetrahedral and the substitutional defect have a $T_d$ point group which is non-centrosymmetric. The hexagonal interstitial on the other hand has a centrosymmetric $D_{3d}$ point group. Unfortunately, we only identified one hexagonal defect $\rm Mn_{i,hex}^0$ combining attractive optical properties and spin multiplicity. However, the excitation energy for $\rm Mn_{i,hex}^0$ is rather small (560~meV) with a transition dipole moment of 2.6~D (see SI Fig. S3). This defect is comparable to the $\rm Se_{Si}^{+}$ defect but with the advantage of centrosymmetry that would increase its optical coherence. Our study indicates that the hexagonal defect configuration is rare with only 60 hexagonal interstitial charged defects in our database. For the majority of the elements and charged states, a defect initialized in an hexagonal position relaxes to a tetrahedral interstitial position. Even when the relaxation keeps the defect hexagonal, the tetrahedral site is still lower in energy for most of the defects (including $\rm Mn_{i,hex}^0$) indicating that the hexagonal configuration is metastable with an energy barrier preventing the relaxation to the tetrahedral site. Only Ni, Ir and B show charged states in which the hexagonal configuration is the ground state interstitial configuration. Our work shows that designing a centrosymmetric defect in silicon emitting in the SWIR would require to turn to complex defects.

\begin{figure}[t]
 	\includegraphics[width=0.5\textwidth]{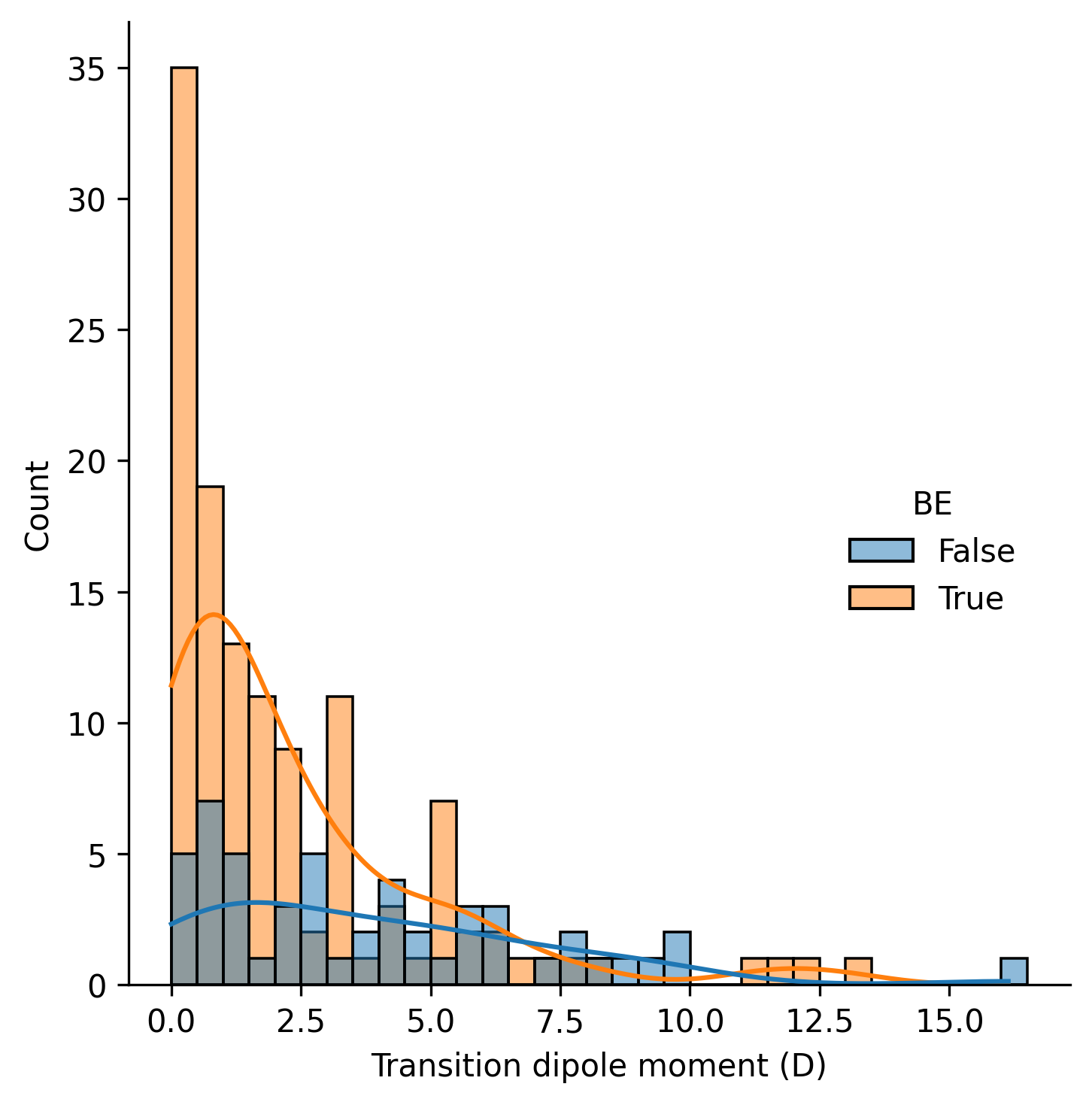}
 	\caption{\label{Fig.4 BE vs. intera defect}Histogram showing the landscape of transition dipole moments of the excitations for all the stable, non-singlet defects. We classified the bound-excitonic and intra-defect transitions using the inverse participation ratio, were the bound-excitonic defects.
}
\end{figure}

A high radiative rate (small radiative lifetime) is also an important technological requirement for a spin-photon interface. Known spin-photon interface candidates in silicon have longer radiative lifetime than their counterparts in diamond. The \textit{T} center and $\rm Se_{Si}^{+}$ show respectively a reported experimental radiative lifetime of 0.94 and 0.9~$\mu$s in agreement with theoretical results obtained with HSE~\cite{Bergeron2020,Dhaliah2022,DeAbreu2019,Xiong2023arXiv}. These values are much larger than the radiative lifetime for the \textit{NV} center in diamond which is in the order of the nanoseconds. Our best defect candidates show also longer radiative lifetime in par with the \textit{T} center and $\rm Se_{Si}^{+}$. This raises the question of how fundamental this higher radiative lifetime is in silicon. The radiative lifetime ($\tau$) depends not only on the TDM (${\boldsymbol{\mu}}$) but also on the frequency of the optical transition ($\nu$) and the refractive index ($n_{r}$): $
{\tau}^{-1}=\frac{n_{r} (2 \pi)^3 \nu^{3}|{\boldsymbol{\mu}}|^{2}}{3 \varepsilon_{0} h c^{3}}$, where $\varepsilon_{0}$ is the vacuum permittivity, $h$ the Planck constant, and $c$ the speed of light. Diamond is the most common host for QIS with the \textit{NV} center being by far the most studied quantum defect. Moving from diamond with a refractive index of 2.4 to silicon with a refractive index of 3.8 increases radiative rate and lowers lifetime. However moving from an optical transition in the visible such as 1.96~eV in \textit{NV} in diamond to the SWIR around 900~meV increases the lifetime (i.e., reduces radiative rate). Both effects taken together increase the radiative lifetime by a factor of 6 for silicon versus diamond for two defects with the same TDM. This is an inherent drawback to working in the SWIR and with silicon. However, a higher TDM could in principle make up for that inherent difference between silicon and diamond. Fig.~\ref{Fig.4 BE vs. intera defect} plots an histogram of the TDM in our database. Very few defects show TDMs higher than the 5 to 6~D of the \textit{NV} center or silicon divacancy in diamond. A transition dipole moment of around 11~D would be needed to make up for the adverse frequency effect to working in the SWIR and reach radiative rates comparable to the \textit{NV} center. Fig.~\ref{Fig.4 BE vs. intera defect} shows that it is very unlikely to attain such a large TDM with simple defects in silicon. Moreover, as shown in Fig.~\ref{Fig.2 TDM_vs_E}a the anticorrelation between optical transition energy and TDM indicates that these high TDM defects are in fact far from being in the SWIR. We hypothesize that the relatively low TDMs are related to the defect-bound exciton nature of many excited states in our dataset. Excitations through bound excitons involve wavefunctions of very different character (localized versus delocalized) leading to low overlap in general~\cite{Xiong2023arXiv}. Fig.~\ref{Fig.4 BE vs. intera defect} supports this analysis by showing a clear difference in TDM distribution for excitation identified as bound exciton. While our data is limited to simple defects, this suggests that defects in silicon might be inherently dimmer than in diamond. While not insurmountable as shown by the recently developed defects in silicon~\cite{Bergeron2020, Higginbottom2022, Morse2017}, this finding has important technological implications. 

While bound excitons lead to lower TDMs and longer radiative lifetime, our study stresses their importance for quantum defects in silicon. All our spin-photon interface candidates are excited through bound excitons when we require optical transitions in the SWIR. The only defect candidate which shows transitions between localized states in the gap is the iodine substitutional $\rm I_{Si}^{2+}$. This defect shows a large transition dipole moment of 5~D but its optical transition at 690 meV is slightly too low to be technologically appealing (see Fig.~S3 in SI). The difficulty in finding defects optically active using only localized defect states without involving host-like states in silicon can be easily rationalized by its small band gap~\cite{Redjem2020}. We note that work from Lee et al.\ surveying certain transition metals in silicon for quantum defects only identified defects with vertical optical excitations lower than 600~meV~\cite{Lee2022}. Our work identifies a series of candidates with higher ZPL energy (from 791 to 939~meV) and longer SWIR wavelength for emission. The discrepancy between the two studies comes from the different criteria used. We included defect candidates that are not triplet (e.g., doublet) and more importantly, we allowed for defects which would be excited through defect-bound excitons. This indicates that relaxing our criteria on the nature of defect states in hosts band gap might be crucial to the discovery of quantum defects in silicon~\cite{Xiong2023arXiv}.

\begin{figure}[t]
 	\includegraphics[width=0.45\textwidth]{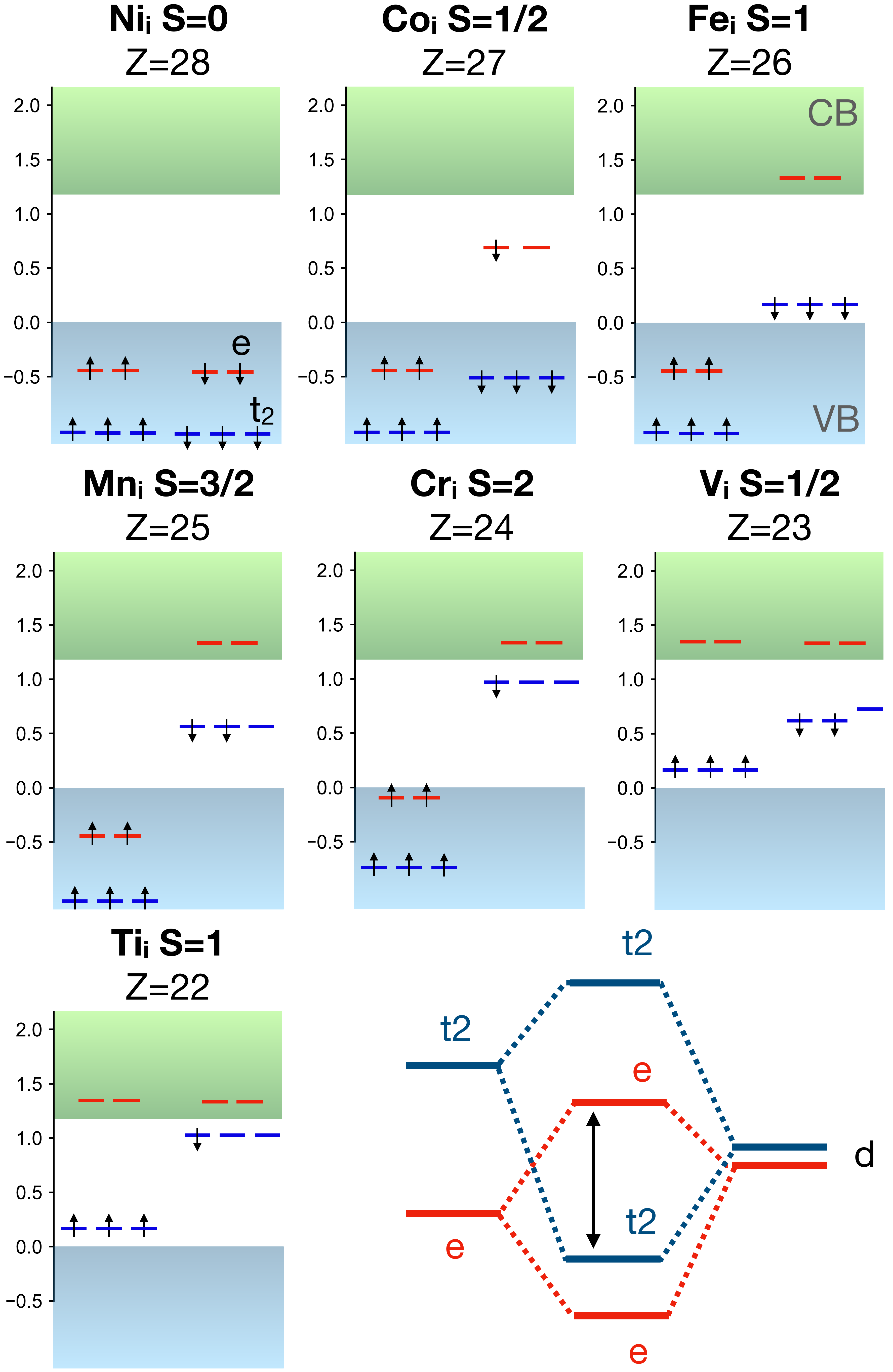}    
 	\caption{\label{Fig.5 3d HSE0}Single-particle levels that are computed at single-shot HSE (HSE$_0$) level for 3$d$ transition metal series in the tetrahedral interstitial configuration for the neutral charge state. The defect levels are represented by the t$_2$ and $e$ symmetry due to the $T_d$ point group.}
\end{figure}

Turning on to the type of chemistry leading to attractive quantum defects, we find that interstitial transition metals are amongst the most interesting with $\rm Ti_{i}^{+}$, $\rm Fe_{i}^{0}$, $\rm Ru_{i}^{0}$ within our top candidates. We can wonder why among all the possibilities, these transition metals and these charge states lead to exceptional properties. Interstitial tetrahedral defects are more favorable energetically than substitutional defects for most of the 3d transition metals except Zn, as shown in Fig.~S5. Both substitutional and interstitial defects are in a $T_d$ point group and form $e$ and $t_2$ defect levels~\cite{Beeler1990, Lindefelt1984, Zunger1982}. The single-particle energy levels for the series of 3$d$ transition metals interstitial defects from Ni to Ti are plotted in Fig.~\ref{Fig.5 3d HSE0} at the single-shot HSE level for the neutral charge state. The raw data is available in Fig.~S7a of SI. A molecular diagram suggested by Beeler et al. is also plotted in Fig.~\ref{Fig.5 3d HSE0} indicating that the $d$ states from the transition metal mixes with the $e$ and $t_2$ state from the silicon host~\cite{Beeler1990}. This simplified molecular diagram does not include spin-splitting. The highest atomic number $Z$ element of the series Ni has entirely filled $t_2$ and $e$ states that are all below the silicon valence band preventing optical activity. As we go to the left of the periodic table (from Ni to Ti) and the atomic number Z decreases, electrons are removed and the atomic $d$ states get higher in energy. The increase in energy of the $d$ states moves the $t_2$ and $e$ levels up and make the $e$ levels move within the band gap. While nickel does not show any possible optical transition to defect levels, cobalt with its $e$ level within the gap can be a potentially optically active defect. When more electrons are removed reaching to $\rm Fe_{i}^{0}$, the $e$ state shifts further and reaches the conduction band while the $t_2$ state moves up in the gap above the valence band. For $\rm Fe_{i}^{0}$, the $t_2$ to $e$ transition is relatively high in energy leading to a technologically attractive ZPL within the near-IR. Moving to Mn and Cr moves the $t_2$ states closer to the middle of the band gap shifting the optical transition to higher wavelength with less appeal for applications. A strong change in total spin happens when moving to V which becomes low spin but still shows excitation energies (i.e., $t_2$ to conduction band) that are on the low side. Finally, Ti leads to the $t_2$ states moving further to the middle of the gap making the lowest energy excitation again too small to be of interest. We can see that for neutral $3d$ transition metal interstitials, 
Fe is in the ``sweet spot'' where the $e$ and $t_2$ levels are close to the valence and conduction bands leading to an optical transition of high energy. While our analysis is based on neutral defects for the sake of simplicity, we can use it to suggest charged defects with appealing electronic structures. Notably, $\rm Ti_{i}^{+}$ which is another of our candidates brings exactly the same filled $t_2$ to $e$ transition as in $\rm Fe_{i}^{0}$ if one electron is removed from the $\rm Ti_{i}^{0}$ defect (see Fig.~\ref{Fig.5 3d HSE0}). Along the same line, other charged defects can be proposed such as $\rm V_{i}^{2+}$ as shown in Fig. S9. However, V is unstable in the $+2$ charge state and was therefore excluded in our screening. To summarize, we rationalize the interest of the Fe$^0$ and $\rm Ti_{i}^{+}$ interstitial defects as they offer among all 3d transition metals a unique position of their $e$ and $t_2$ states leading to high excitation energy and SWIR emission. 

\begin{figure}[t]
 	\includegraphics[width=0.45\textwidth]{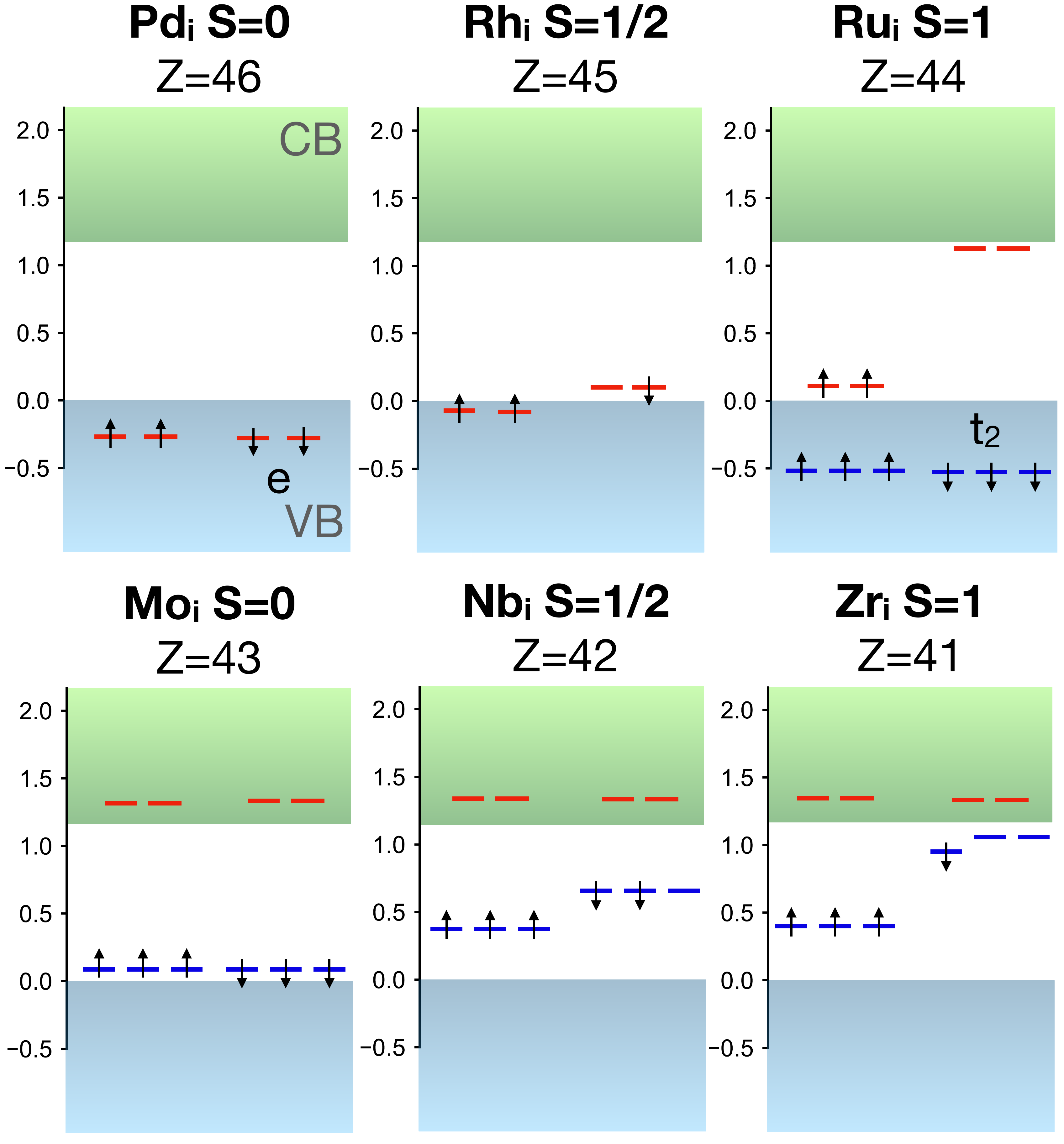}    
 	\caption{\label{Fig.6 4d HSE0}Single-particle levels that are computed at single-shot HSE (HSE$_0$) level for 4$d$ transition metal series in the tetrahedral interstitial configuration with neutral charges.}
\end{figure}

A similar analysis can be performed for the 4$d$ series (Pd to Zr) as illustrated in Fig.~\ref{Fig.6 4d HSE0} (cf.\ Fig.~S7b for the raw data). There, we see that the $e$ to conduction band transition suggested for cobalt in $\rm Co_{i}^{+}$ can be mimicked for $\rm Ru_{i}^{0}$. The $t_2$ to $e$ transition found in $\rm Fe_{i}^{0}$ is on the other hand present in the neutral $\rm Mo_{i}^{0}$. $\rm Mo_{i}^{0}$ ground state is a singlet ($S=0$) but shows an attractive excitation energy (0.99 eV) with a bright emission (1.92D) at the HSE level (Fig.~S8d). This defect could still be of use as single-photon emitter but not as spin-photon interface. While not the focus of the present study, $\rm Mo_{i}^{0}$ is probably worth experimental investigation as it shows properties in par with the \textit{G} center, one of the most interesting singlet quantum defect in silicon~\cite{Komza2022arXiv,Redjem2020,Prabhu12022arXiv}. As in the 3$d$ series, the removal of electrons in neutral 4$d$ defects can lead to charged defects with the same $t_2$ to $e$ transition found in $\rm Fe_{i}^{0}$ or $\rm Ti_{i}^{0}$ -- not surprisingly we find this is the case for one of our candidates $\rm Zr_{i}^{+}$, however its ZPL is on the low side (666~meV). This would also be the case for a hypothetical $\rm Nb_{i}^{2+}$ which similar to $\rm V_{i}^{2+}$ shows a very attractive electronic structure but is unstable (see SI, Fig. S8a).  When considering 5$d$ elements, we find a similar electronic structure for $\rm Hf_{i}^{+}$ as $\rm Zr_{i}^{+}$ but with even lower optical transition energies. For the later 5$d$ transition metals in our data set (i.e., Os and Ir), the presence of a 6$s$ state in the band gap prevents attractive first excitation as the $d$ to $s$ transition are forbidden (Fig.~S8d of SI). The second excitation is however allowed and in an interesting energy range. For instance, $\rm Os_{i}^{0}$ shows a second excitation at 0.93 eV just above the dipole forbidden 0.82 eV excitation at the full HSE level. So, this emitter could operate but with a dark state that might be detrimental to efficiency. Additionally, as for $\rm Mo_{i}^{0}$ in the singlet 4ds, $\rm W_{i}^{0}$ is an attractive singlet in the 5ds (Fig.~S8d of SI).

The substitutional transition metals can be similarly analyzed. The substitutional defects are also in a tetrahedral $T_d$ point group forming $t_2$ and $e$ states but contrary to the interstitial case, the $e$ state is lower in energy than the $t_2$ state~\cite{Beeler1990, Lindefelt1984, Zunger1982}. The $t_2$ state comes from mixing between the $t_2$ $d$ states from the transition metals and the $t_2$ from the vacancy. The $e$ state is not as bonding as in the interstitial case~\cite{Beeler1990}. We report the electronic structures in the substitutional state for 3$d$ metals in silicon in the SI (Fig.~S7c of SI). Contrary to the interstitial, the $t_2$ states are present in the band gap in Ni. As we move towards Ti,  the number of electrons decreases and $Z$ increases, moving the $t_2$ states higher in energy. Remarkably the $e$ states are never in the gap and are either below the valence band or above the conduction band except for Mn. Only the higher atomic number transition metals such as Ni and Co (more specifically $\rm Ni_{Si}^{-}$ and $\rm Co_{Si}^{2-}$) offer possibilities for optical transitions between the $t_2$ state and conduction band-like states. Any substitutional 3$d$ transition metal beyond Co (from Fe to Ti) does not show any single-energy particle state within the band gap making them not attractive for optical emission except Mn$_{\rm Si}^0$. It is only when we move to 4$d$ and 5$d$ defects that we see earlier transition metals such as $\rm Nb_{Si}^{-}$ and $\rm Ta_{Si}^{-}$ becoming potentially optically active in the SWIR. For instance, compared to V which is the 3$d$ in the same column, the higher energy of the 4$d$ and 5$d$ orbitals in Nb and Ta, lead to a higher $e$ states that is now in the gap and can lead to potential excitonic transitions (see Fig. S7). Remarkably, while the substitutional transition metals can have attractive energy levels well positioned with the gap, they appear for negatively charged defects that are unstable as donor bound excitons. The electron count in the interstitial case on the contrary can lead to adequate excitation in neutral or positive charge states which favor stability of the donor bound exciton. The only non-transition metal selected in our shortlist of 7 potential candidates is the $\rm Na_{Si}^{0}$ (see Table I). This defect was selected because of a possible transition between 3 degenerate defect states to the conduction band as shown in Fig.~S3. The unstable bound exciton excluded this defect for further consideration but it is remarkable that these 3 defect states are not related to any the alkali levels. The closest alkali levels are just below the valence band. The three defect states are localized on the silicon neighbors to the sodium substitutional. We relate these defect states to the strain induced by the alkali. This effect is observed with potassium and rubidium as well (see Fig.~S3 of SI).

Remarkably, we did not find any interesting quantum defect candidates from main group elements. We recovered the previously studied $\rm Se_{Si}^{+}$ but it was excluded in our screening because of its inappropriate emission wavelength~\cite{Morse2017,DeAbreu2019}. Here, the selenium substitutional defect can be seen as a Se$^{6+}$ substituting on Si$^{4+}$ with one electron left. We note that Abraham et al.\ suggested that reproducing this electronic configuration could be considered using interstitial alkali-earth that have a $+2$ oxidation state and could be made in a multiplet $+1$ charged state~\cite{Abraham2018}. The experimental study of from Abraham et al.\ on Mg interstitials in silicon however did not show any PL. Our database indicates that the alkali-earth series Be, Mg, Ca, Sr, Ba in interstitial position do not show levels that are adequately positioned with respect to the band edges (see Fig.~S9 of SI). The best contender is $\rm Be_{Si}^{+}$ but it does not show strong TDM (much lower than $\rm Se_{Si}^{+}$) and a transition energy that is still too far in the infra-red. This last example on alkali-earth interstitial illustrates how access to a database of computed defect properties can help identify or rule out proposed defects following certain design guidelines.

\section{Conclusion}

While silicon has become an emerging host for optically controlled quantum defects, an optimal color center to use as a spin-photon interface is still under active search. We have used high-throughput computational screening to identify a handful defect candidates among a database of more than 1000 simple charged defects in silicon. The emergence of specific tetrahedral interstitial transition metals (i.e., $\rm  Fe_{\rm i}^0$, $\rm Ru_{\rm i}^0$, $\rm  Ti_{\rm i}^+$) was rationalized through the positions of their defect induced $t_2$ and $e$ states.  The use of a single-shot HSE$_0$ approach was critical in scaling up our screening to thousands of charged defects without compromising the overall accuracy. We found that the defects offering strong optical emission at a technologically relevant wavelength are rare and involve states close to the band edges forming defect-bound excitons. The excitonic nature of these defects and our dataset suggest that silicon defects might be intrinsically dimmer than defects in diamond. Our work motivates further experimental and theoretical investigations of the unique identified defect candidates and outlines an effective high-throughput approach to future screening for quantum defects including complexes in silicon and other hosts.

\section{Methods}

We considered 56 elements that are ion-implantable, as highlighted in the periodic table in Fig.~S5. This collection covers the majority of the elements except the rare-earth, the noble gas, and some of the 5$d$ elements. All defect computations at DFT level were performed using the automatic workflows that are implemented in atomate software package~\cite{Jain2013, Mathew2017,Ong2013}. The DFT calculations were performed using Vienna Ab-initio Simulation Package (VASP)~\cite{G.Kresse-PRB96,G.Kresse-CMS96} and the projector-augmented wave (PAW) method~\cite{P.E.Blochl-PRB94} with the Perdew-Burke-Erzhenhoff (PBE) functional. To simulate charged defects, we applied the spin-polarized computations and used a supercell size of 216 atoms. 520 eV cutoff energies were used for the plane-wave basis and only the $\Gamma$ point was used to sample the Brillouin zone. The defect structures were optimized at a fixed volume until the forces on the ions are smaller than 0.01~eV/$\AA$. 
After these DFT computations, we applied the Heyd-Scuseria-Ernzerhoff (HSE)~\cite{Heyd2003} hybrid-functional calculations but in a ``single-shot'' fashion (HSE$_0$) to all the defects in the database for improved descriptions of single-particle levels at a significantly reduced computational cost compare to the full HSE. Using the defect structures in 216 atoms that are optimized at DFT level, we carried out the perturbative single-shot approach in which the single-particle eigenvalues are calculated using the PBE wavefunctions. We used the standard 25\% Fock exchange in these single-shot calculations. We note that the HSE$_0$ band gap of Si is 1.18 eV, in good agreement with full HSE results. Here we report the relevant INCAR tags to perform the single-shot HSE$_0$ calculation: $\rm ALGO = Eigenval$, $\rm LHFCALC = True$, $\rm HFSCREEN = 0.2$, $\rm AEXX = 0.25$, $\rm ICHARG = 1$, $\rm NELM = 1$.

For the 19 screened quantum defect candidates, we applied the fully self-consistent HSE hybrid functional with 25\% exact exchange. We further increased the supercell size to 512 atoms, with a reduced plane-wave basis cutoff energies of 400~eV and a $\Gamma$ only k-point. We performed the relaxations using a force criteria of 0.01~eV/$\AA$. The HSE functional provides an accurate band gap of 1.114~eV for silicon, which is in excellent agreement with the measured band gap of 1.17~eV at 0~K. For all defect computations symmetry was not imposed.

The defect formation energies were analyzed using PyCDT~\cite{Broberg2018}. We computed the formation energy of each charged-defect state as a function of the Fermi level $E_f$~\cite{Zhang1991,Komsa2012}:
\begin{equation}
    E_\mathrm{form}[X^q] = E_\mathrm{tot}[X^q] - E_\mathrm{tot}^\mathrm{bulk} - \sum n_i \mu_i + q E_f + E_\mathrm{corr}
\label{eq:defect}
\end{equation}
where $E_\mathrm{tot}[X^q]$ and $E_\mathrm{tot}^\mathrm{bulk}$ are the total energies of the defect-containing supercell (for a given defect $X$ in the charge state $q$) and the bulk, respectively. The third term represents the energy needed to exchange atoms with thermodynamic reservoirs where $n_i$ indicates the number of atoms of species $i$ removed or added to create the defect, and $\mu_i$ their corresponding chemical potential. The fourth term represents the energy to exchange electrons with the host material through the electronic chemical potential given by the Fermi level. Finally, the last term is the finite-size correction accounting for the spurious Coulomb interactions under periodic boundary conditions~\cite{Freysoldt2011,Kumagai2014}.

The zero-phonon line (ZPL) energy is defined as the difference between the energy of the excited state and that of the ground state. We obtained the total energy of the excited states for the relaxed defects through the constrained-HSE method that forces the occupation of the unoccupied single-particle states. This methodology has been shown to give ZPLs within 100~meV from experiment for defects in diamond and silicon~\cite{Gali-PRL2008, S.Li-naturecomm2022, Xiong2023arXiv}. When searching for most relevant excited states, we consider the transitions that gives the smallest energy difference, while also allowing an energy window of up to 100~meV to take into account the errors and band degeneracy. For all the possible transitions that identified in the described process, we further compute the transition dipole moment, and the pair of initial and final bands that give the largest transition dipole moment (TDM) are identified as the most relevant defects. The TDM measures the tendency of the transition between the initial and final states, and it was calculated using using the wavefunctions from VASP and analyzed by the PyVaspwfc package \cite{Zheng2018}. For a single $k$ point, the TDM is defined as:

\begin{equation}
\boldsymbol{\mu}_{k}=\frac{\mathrm{i} \hbar}{\left(\epsilon_{\mathrm{f}, k}-\epsilon_{\mathrm{i}, k}\right) m}\left\langle\psi_{\mathrm{f}, k}|\mathbf{p}| \psi_{\mathrm{i}, k}\right\rangle,
\end{equation}
where $\epsilon_{\mathrm{i}, k}$ and $\epsilon_{\mathrm{f}, k}$ stands for the eigenvalues of the initial and final states, $m$ is electron mass, $\psi_{\mathrm{i}}$ and $\psi_{\mathrm{i}}$ stands for the initial and final wavefunctions, and $\mathbf{p}$ is the momentum operator. We note that the optical transition dipole moment can also be computed through from the derivative of the wavefucntions which is provided in the vasp optical calculations. We carefully examine few systems and compared the results of pyvaspwfc and from WAVEDER from vasp. The results show excellent correspondence between these methods. The detailed benchmark is given in Fig.~S10 of SI.

PL spectra were computed following the procedure of Alkauskas et al.~\cite{Alkauskas2014} with a mixed GGA+HSE approach. In this mixed approach, the ground and excited state structures are fully relaxed at the HSE level, while the phonons are calculated within GGA for a carefully optimized GGA groundstate structure. This approach has been successfully used for the \textit{NV} center providing excellent agreement between the computed and experimental Huang-Rhys factor~\cite{Alkauskas2014}. Using only GGA for both structure relaxations and phonon calculations has been shown to underestimate the experimental Huang-Rhys factor of the NV center by about 20\% ~\cite{Razinkovas2021}. Here, we found that the values for the Huang-Rhys factors with GGA only are similar to those of the mixed GGA+HSE approach reported in the Results section. Namely, the GGA only Huang-Rhys factors are 0.15, 0.12, and 0.27 for $\rm Ti_{i}^{+}$, $\rm Fe_{i}^{0}$, and $\rm Ru_{i}^{0}$, respectively. The defect structures were simulated in a 512-atoms cell. This supercell size has been demonstrated to yield good convergence of the Huang-Rhys factor of the NV center ~\cite{Alkauskas2014,Razinkovas2021}. The phonons of the ground and excited states were assumed to be identical within the so-called equal-mode approximation~\cite{Razinkovas2021}. 
The phonons (phonon-frequencies and eigenvectors) were calculated within the finite-displacement method in the harmonic approximation using the \texttt{Phonopy} package~\cite{Togo2015}. To this end, the ground state structure was first carefully relaxed with GGA until forces were smaller than 0.001 eV/\AA. The atoms in the relaxed structure were then systematically displaced (atomic displacement magnitude was set to $0.01\,$\AA) and the restoring forces were computed with VASP. The calculated forces determine the atomic force constants from which the dynamical (Hessian) matrix is formed. Diagonalizing the dynamical matrix yields the phonon eigenfrequencies and eigenvectors that are key ingredients to calculating the phonon sidebands of the PL spectra~\cite{Alkauskas2014}. A Gaussian broadening of 2 meV was used for the PL spectra plots in Fig. S6. The ZPL positions were set to the values reported in table~\ref{Tab: non-singlet defects}.

\section{Competing interests}
The authors declare no competing financial interest.

\section{Contributions}
Y. X. and D. D. developed the high-throughput workflow and performed the computations. Y. X. and C. B. performed the HSE computations. N. S. performed the PL computations. Y. X., C. B., W. C., G.H., A. S., S. M. G. and H. S. analyzed the data. S. M. G., A. S., and G. H. supervised and designed the project. The manuscript was drafted and edited by all the authors at each stage of its preparation.

\section{Acknowledgments}
\begin{acknowledgments}
This work was supported by the U.S. Department of Energy, Office of Science, Basic Energy Sciences 
in Quantum Information Science under Award Number DE-SC0022289. 
This research used resources of the National Energy Research Scientific Computing Center, 
a DOE Office of Science User Facility supported by the Office of Science of the U.S.\ Department of Energy 
under Contract No.\ DE-AC02-05CH11231 using NERSC award BES-ERCAP0020966. 
\end{acknowledgments}

\end{document}